\documentclass[a4paper,11pt]{article}
\pdfoutput=1 

\usepackage{jcappub} 

\usepackage[T1]{fontenc} 
\usepackage{graphicx}
\usepackage{epsfig}
\usepackage{rotate}
\usepackage{amsmath}
\usepackage{amssymb}
\usepackage{amsfonts}
\usepackage{bm}
\usepackage{tablefootnote}
\usepackage{enumerate}
\usepackage{afterpage}
\usepackage{xcolor}
\usepackage{natbib, hyperref}

\usepackage{multicol}
\usepackage{multirow}
\usepackage{booktabs}

\UseRawInputEncoding

\allowdisplaybreaks

\def\be{\begin{equation}}
\def\ee{\end{equation}}
\def\bea{\begin{eqnarray}}
\def\eea{\end{eqnarray}}

\newcommand{\blue}[1]{{\color{black}{#1}}}

\newcommand{\orcid}[1]{\href{https://orcid.org/#1}{\,\includegraphics[width=8px]{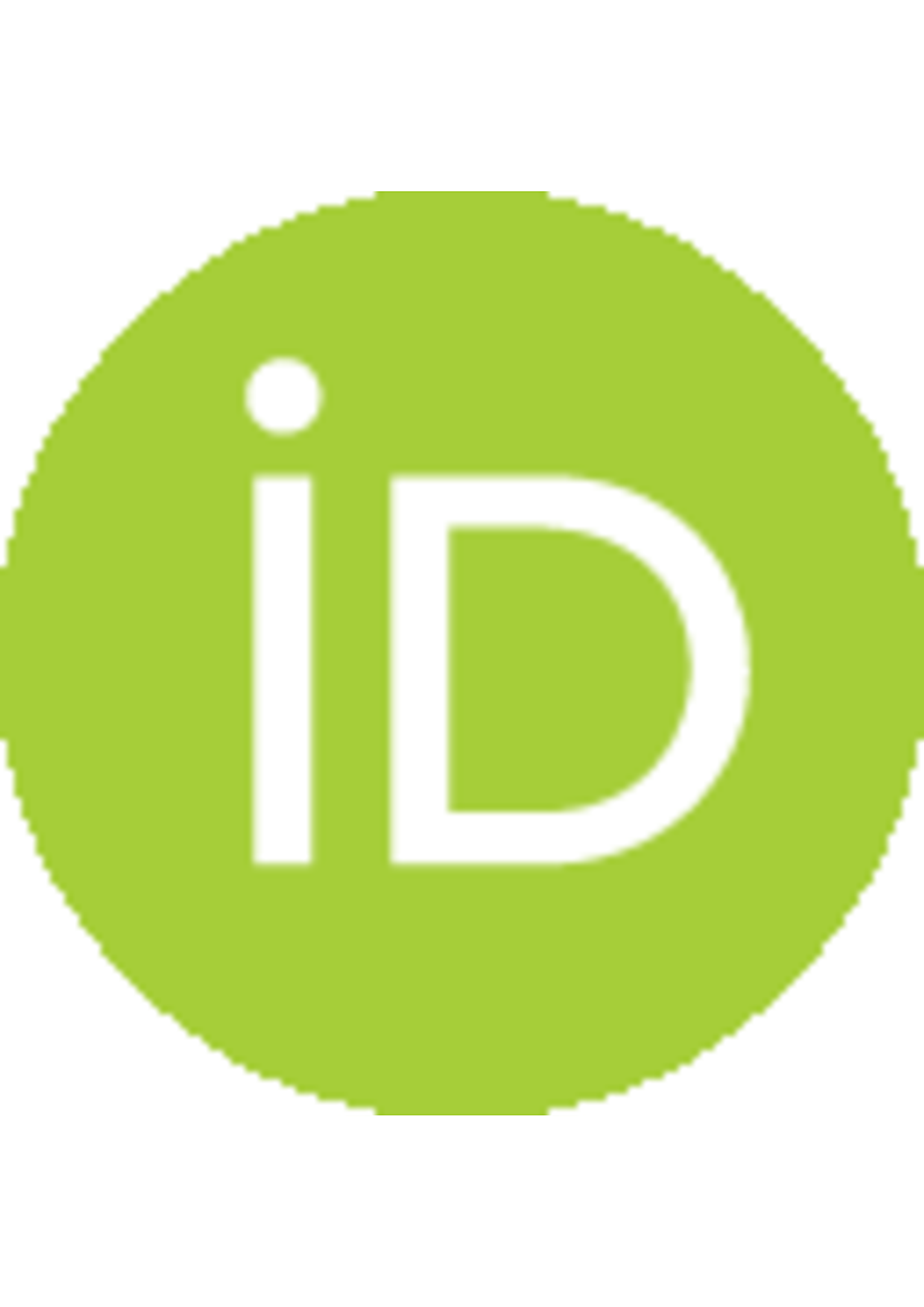}}}

\usepackage{pgf,tikz}
\usetikzlibrary{arrows}

\usepackage{soul}
\usepackage{xcolor}
\setstcolor{blue}
\usepackage{cancel}

\title{{Phantom crossing or dark interaction?}}

\author{
S\^ecloka L. Guedezounme$^{a}$\orcid{0009-0000-9200-8584},
Bikash R. Dinda$^{a,b}$\orcid{0000-0001-5432-667X},\\
Roy Maartens$^{a,b}$\orcid{0000-0001-9050-5894}
}

\affiliation[a]{Department of Physics \& Astronomy, University of the Western Cape, Cape Town 7535, South Africa}

\affiliation[b]{National Institute for Theoretical \& Computational Science, Cape Town 7535, South Africa}

\emailAdd{seclokaguedezounme@gmail.com}
\emailAdd{bikashrdinda@gmail.com}
\emailAdd{rmaartens@uwc.ac.za}

\abstract{
Recent results from DESI BAO measurements, together with Planck CMB and Pantheon+ data, suggest that there may be a `phantom' phase ($w_{\rm de}<-1$) in the expansion of the Universe. This inference follows when the $w_0, w_a$ parametrization for the dark energy equation of state $w_{\rm de}$ is used to fit the data. Since phantom dark energy in general relativity is unphysical, we investigate the possibility that the phantom behaviour is not intrinsic, but effective -- due to a non-gravitational interaction between dark matter and non-phantom dark energy. To this end, we assume a physically motivated {thawing quintessence-like} form of the intrinsic dark energy equation of state $w_{\rm de}$. Then we use a $w_0, w_a$ model for the {\em effective} equation of state of dark energy.
We find that the data favours a phantom crossing for the effective dark energy, but only at low significance.
The intrinsic equation of state of dark energy is non-phantom, without imposing any {non-phantom} priors. A nonzero interaction is favoured at more than $3\sigma$ {at $z\sim0.3$}. The energy flows from dark matter to dark energy at early times and reverses at later times.
}

\usepackage[toc]{appendix}

\begin{document}

\maketitle
\date{\today}


\section{Introduction}
\label{sec1}

The origin and dynamics of dark matter and dark energy, which together account for approximately 95\% of the total current energy content of the Universe, remain among the greatest unsolved questions in modern cosmology. In the standard $\Lambda$CDM model, dark energy is modelled as a cosmological constant $\Lambda$, and dark matter is treated as a cold, pressureless component, with no non-gravitational interaction between them \cite{Frieman:2008sn, Bertone:2004pz,Planck:2018vyg}.

While this framework provides an excellent fit to a wide range of cosmological observations, it faces several theoretical and empirical challenges, including tensions in measurements of the Hubble constant and large-scale structure growth, the cosmic coincidence problem, and the cosmological constant problem (see e.g. \cite{Zlatev:1998tr,Weinberg:2000yb,Hildebrandt:2016iqg,Verde:2019ivm,DES:2021wwk,delaMacorra:2021hoh, DiValentino:2021izs,Yashiki:2025loj,Dinda:2021ffa,Brax:2025ahm,Nojiri:2025low,RoyChoudhury:2024wri,RoyChoudhury:2025dhe,Sabogal:2025jbo,Silva:2025twg,Cheng:2025lod,Afroz:2025iwo,Mishra:2025goj,Banerjee:2020xcn}). These issues motivate the exploration of alternative models, including those in which dark matter and dark energy interact non-gravitationally 
(see e.g. \cite{Zimdahl:2012mj,Pourtsidou:2013nha,Wang:2016lxa,Montani:2025rcy,Khoury:2025txd,Li:2024qso,Li:2025owk,Dinda:2022ixi,Valiviita:2009nu,Clemson:2011an,Yan:2025iga,Guin:2025xki,Cruz:2019kgz,Cardenas:2018nem,Shah:2025ayl,Bedroya:2025fwh}).

These models introduce an interaction term $Q(z)$ in the background energy conservation equations, allowing for energy exchange between the dark components. The sign of $Q$ determines the direction of energy flow: $Q>0$ corresponds to energy transfer from dark energy to dark matter, and vice-versa for $Q<0$. Several studies have shown that dark interaction models can address some of the potential issues in cosmology, such as the Hubble tension and the crossing of the phantom divide   \cite{Yang:2015tzc, DESI:2025fii, DESI:2025zgx, Yang:2025boq, Zhai:2025hfi}.

However, a major challenge in analysing dark interaction models arises from the degeneracy between the interaction $Q(z)$ and the dark energy equation of state parameter $w_{\rm de}(z)$. Both quantities influence the expansion history of the Universe, making it impossible to  {study their effects individually}, using background observations alone, without additional assumptions \cite{Yang:2015tzc, Escamilla:2023shf,Dinda:2024ktd, Zhai:2025hfi, Abedin:2025yru, Li:2025ula, Chakraborty:2025syu, Linder:2025zxb}. 
Model-dependent approaches assume specific functional forms for the interaction, but typically these are not physically well-motivated (see e.g. \cite{Pourtsidou:2013nha}). Here we do not make assumptions about $Q(z)$ -- instead, we adopt a physically motivated form for the {intrinsic} $w_{\rm de}(z)$ and the $w_0w_a$ parametrization for the effective equation of state of dark energy $w_{\rm de}^{\rm eff}(z)$, to break the degeneracy between $w_{\rm de}(z)$ and  $Q(z)$ (see e.g. \cite{Escamilla:2023shf, Abedin:2025yru, Li:2025ula} for related work). 

Recent advances in observational cosmology, especially from the Dark Energy Spectroscopic Instrument Data Release 2 (DESI DR2)  baryon acoustic oscillation (BAO) measurements \cite{DESI:2025fii, DESI:2025zgx}, the Planck cosmic microwave background (CMB) data \cite{aghanim2020planck}, and the Pantheon+ supernova compilation \cite{Brout:2022vxf}, provide high precision for probing the dark sector and we study the dark interaction in light of these observations. 

{The paper is organised as follows. In \autoref{sec:basics}, we present the theoretical framework for the interacting dark sector and derive the key equations governing the interaction. 
The observational data are briefly reviewed in \autoref{sec:data} and the results are presented in \autoref{sec:results}. We conclude in \autoref{sec:conclusion} with a discussion of the implications of our findings.}


\section{Background dynamics of dark interaction}
\label{sec:basics}

We consider a spatially flat Friedmann-Lema\^itre-Robertson-Walker (FLRW) universe governed by general relativity, where the dark sector consists of dark matter and dark energy with equation of state $w_{\rm de} = p_{\rm de}/\rho_{\rm de}$. The total energy density satisfies the Friedmann equation:
\begin{align}
    H(z)^2 = \frac{8\pi G}{3} \big[ \rho_{\rm dm}(z) + \rho_{\rm de}(z) + \rho_{\rm b}(z) + \rho_{\rm r}(z) \big] \, , \label{eq:Hsqr}
\end{align}
where $\rho_{\rm a}(z)$ denotes the energy densities of dark matter, dark energy, baryons, and radiation, respectively, and $H(z)$ is the Hubble parameter.

In the presence of a non-gravitational interaction between dark matter and dark energy, the standard conservation equations are modified:
\begin{align}
    \dot{\rho}_{\rm dm} + 3H\rho_{\rm dm} &= Q \, ,
    \label{eq:dm_conservation}
    \\
    \dot{\rho}_{\rm de} + 3H\big(1 + w_{\rm de}\big)\rho_{\rm de} &= -Q \, . \label{eq:de_conservation}
\end{align}
Radiation and baryons are assumed to be separately conserved at late times:
\begin{align}
    \dot{\rho}_{\rm b} + 3H\rho_{\rm b} = 0, \quad \dot{\rho}_{\rm r} + 4H\rho_{\rm r} = 0 \, . \label{eq:b_r_conservation}
\end{align}

\autoref{eq:dm_conservation} and \autoref{eq:de_conservation} can be equivalently described by a non-interacting model with effective equations of state: 
\begin{align}
\label{eq:wdm_eff}
    w^{\rm eff}_{\rm dm}(z) &= -\frac{Q(z)}{3H(z)\rho_{\rm dm}(z)}\, , \\
\label{eq:wdeff}
    w^{\rm eff}_{\rm de}(z) &= w_{\rm de}(z) + \frac{Q(z)}{3H(z)\rho_{\rm de}(z)}\, ,
\end{align}
corresponding to
\begin{align}
\label{eq:eff_dm_consrvn}
	\dot{\rho}_{\rm dm} + 3H \big(1+w^{\mathrm{eff}}_{\rm dm}\big) \rho_{\rm dm} &= 0 \, , \\
\label{eq:eff_cnsrvn_de}
    \dot{\rho}_{\rm de} + 3H \big(1+w^{\mathrm{eff}}_{\rm de}\big) \rho_{\rm de} &= 0 \, .
\end{align}
{\autoref{eq:wdm_eff} and \autoref{eq:wdeff} show that among the four quantities $w_{\rm de}$, $w_{\rm de}^{\rm eff}$, $Q$, and $w_{\rm dm}^{\rm eff}$, only two are independent. Hence the interaction model can be fully specified by any two of these quantities. Here we specify the model through $w_{\rm de}$ and $w_{\rm de}^{\rm eff}$, which fixes  $Q$ and $w_{\rm dm}^{\rm eff}$.}

\autoref{eq:Hsqr} {is} rewritten as
\begin{equation}
\label{eq:H_background}
    H(z) = H_0 \left[
        \Omega_{\rm r,0}(1+z)^4 +
        \Omega_{\rm b,0}(1+z)^3 +
        \Omega_{{\rm dm},0} f_{\rm dm}(z) +
        \Omega_{{\rm de},0} f_{\rm de}(z)
    \right]^{1/2},
\end{equation}
where $\Omega_{\rm a,0} = \rho_{\rm a,0} / \rho_{\rm crit,0}$ and  $\rho_{\rm crit,0} = 3H_0^2 / (8\pi G)$. The normalized energy densities of dark matter and dark energy are
$f_{\rm dm}(z) = \rho_{\rm dm}(z) / \rho_{\rm dm,0}$ and $f_{\rm de}(z) = \rho_{\rm de}(z) / \rho_{\rm de,0}$ and 
 $\Omega_{\rm r,0}+\Omega_{\rm b,0}+\Omega_{{\rm dm},0}+\Omega_{{\rm de},0}=1$. In the standard non-interacting case $Q = 0$, we have $f_{\rm dm} = (1 + z)^{3}$, $f_{\rm de} = \exp{\left[3 \, \int_{0}^{z} \,\mathrm{d} z' [1 + w_{de}(z')]/( 1 + z') \right]}$.
The present-day radiation density parameter is 
\begin{align}
    \Omega_{\rm r,0} = \frac{\Omega_{\rm m,0}}{1 + z_{\rm eq}} ,
\end{align}
where the matter-radiation equality redshift in standard early-time physics is well approximated as \cite{Chen:2018dbv}
\begin{align}
    z_{\rm eq} \approx 2.5 \times 10^4 \, \Omega_{\rm m,0} h^2 \left( \frac{T_{\rm CMB}}{2.7\,\mathrm{K}} \right)^{-4}.
\end{align}
Here $h = H_0 / (100\,\mathrm{km\,s^{-1}Mpc^{-1}})$ and $T_{\rm CMB} \approx 2.7255$ K is the background CMB temperature.


The degeneracy between $Q(z)$ and $w_{\rm de}(z)$ means that it is not possible, from background data, to reconstruct either of these independently without assuming any model or parametrization. We aim to reconstruct $Q(z)$  by using a physically motivated parametrization of $w_{\rm de}(z)$, corresponding to thawing quintessence models. In these models, a minimally coupled scalar field is frozen at early times due to Hubble friction, effectively behaving as a cosmological constant with $w_{\rm de} \approx -1$ \cite{Gialamas:2025pwv}. It evolves later, leading to a deviation from a cosmological constant at late times with increasing $w_{\rm de}>-1$. We adopt the analytic from \cite{DESI:2025fii}:
\begin{align}
    w_{\rm de}(z) = -1 + (1 + w_{\rm q}) (1+z)^{-1} \left[\frac{3}{2 +  (1 + z)^{3}}\right]^{2/3}\quad\mbox{where} \quad  w_{\rm q}=w_{\rm de}(0)\,. \label{eq:w_de_thawing}
\end{align}
Note that the above form corresponds to thawing quintessence only when $w_{\rm q}>-1$. On the other hand, it behaves like phantom dark energy for $w_{\rm q}<-1$. This is the intrinsic dark energy equation state. However, observations will measure the effective equation of state, for which we assume 
the standard $w_0,w_a$ parametrization 
\begin{align}
    w^{\rm eff}_{\rm de}(z) = w_0 + w_a \frac{z}{1 + z} \, . \label{eq:w_eff_CPL}
\end{align}
Then $Q$ is determined by $w_{\rm de}$ and $w_{\rm de}^{\rm eff}$. 
Our aim now is as follows:
\begin{itemize}
    \item 
 to show that the data allows for an interaction with intrinsic and effective equations of state,  of the forms \autoref{eq:w_de_thawing} and \autoref{eq:w_eff_CPL};
    \item
 to then find the level of interaction required for this to be the case.
\end{itemize}

From \autoref{eq:eff_cnsrvn_de}, we have
\begin{align}
f_{\rm de}(z) = 
\exp\left\{3\!\int_{0}^{z}  \left[1 + w^{\mathrm{eff}}_{\rm de}(z')\right]   \frac{\mathrm{d} z'}{1+z'} \right\} = (1+z)^{3(1+ w_{0} + w_{a})} \,   \exp{\left(\!- 3 w_{a}\, \frac{z}{1+z} \right)} \, .
\end{align}
Then \autoref{eq:wdeff} gives
\begin{equation}
    Q(z) = 3 H(z) \rho_{\rm crit,0}\, \Omega_{\rm de,0} \big[w^{\rm eff}_{\rm de}(z)-w_{\rm de}(z)\big] f_{\rm de}(z) \,,
    \label{eq:Q_case1}
\end{equation}
and it follows from \autoref{eq:wdm_eff} that
\begin{equation}
    w^{\rm eff}_{\rm dm}(z) = 
    \big[w_{\rm de}(z) - w^{\rm eff}_{\rm de}(z)\big]\,
    \frac{  \Omega_{\rm de,0}\, f_{\rm de}(z)}{ \Omega_{\rm dm,0}\, f_{\rm dm}(z)} \, .
    \label{eq:wdm_eff_case1}
\end{equation}
Substituting into \autoref{eq:eff_dm_consrvn}, we find a differential equation for $f_{\rm dm}$:
\begin{equation}
    \dfrac{{\rm d}f_{\rm dm}}{{\rm d}z} = \frac{3}{1+z} \left[ f_{\rm dm} +
    \big(w_{\rm de}-w^{\rm eff}_{\rm de} \big) \,
    \frac{\Omega_{\rm de,0}\, f_{\rm de}}{\Omega_{\rm dm,0}} \right] \, ,
    \label{eq:fdm_eqn_case1}
\end{equation}
with solution 
\begin{align}
  f_{\rm dm}(z) &= (1 + z)^3 \Bigg\{1 +  3\frac{\Omega_{\rm de, 0}}{\Omega_{\rm dm, 0}} \, \int_{1}^{(1+z)^{-1}} \, \mathrm{d}a  \bigg[\left(1 + w_{0} + w_{a}\right) - w_{a} a 
  \\\notag
  & \qquad\qquad\quad~ - (1 + w_{\rm q})\, a^2 \left( \frac{3}{2a^3 + 1} \right)^{2/3} \bigg] \, {a}^{-(1 + 3 w_{0} + 3 w_{a})} \,   \exp{\big[- 3 w_{a} (1-a) \big]}   \Bigg\} \, .
  \label{eq:fdmexp} 
\end{align}
We can then infer the Hubble parameter from \autoref{eq:H_background} and constrain the parameters $w_{\rm q}$, $w_{0}$, $w_{a}$ and other cosmological parameters.

\blue{The effective equation of state of all the components of the model can be approximated as}
\begin{equation}
\blue{
w_{\rm tot}^{\rm eff}(z) = E^{-2} (z) \left[ w^{\rm eff}_{\rm de}(z) \, \Omega_{{\rm de},0} \, f_{\rm de}(z) + w^{\rm eff}_{\rm dm}(z) \, \Omega_{{\rm dm},0} \, f_{\rm dm}(z) + \frac{1}{3} \Omega_{\rm r,0}(1+z)^4 \right] \, ,
}
\label{eq:effective_w_tot}
\end{equation}
\blue{where $E(z) = H(z)/H_0$. At late times, we can neglect the radiation components and hence the last term in this equation.}


\section{Observational data}
\label{sec:data}

In order to constrain parameters and reconstruct the dark sector interaction function $Q(z)$, we use three complementary cosmological datasets:

\begin{itemize}
    \item \textbf{DESI DR2 BAO} \cite{DESI:2025zgx}: Measurements of BAO from DESI DR2 provide high-precision determinations of the comoving angular diameter distance $D_M(z)$, the Hubble distance $D_H(z) = c/H(z)$, and the volume-averaged distance $D_V(z)$. These are reported in units of the comoving sound horizon at the drag epoch $r_d$ (i.e. uncalibrated):
    \begin{align}
        \widetilde{D}_M(z) = \frac{D_M(z)}{r_d}, \qquad
        \widetilde{D}_H(z) = \frac{D_H(z)}{r_d}, \qquad
        \widetilde{D}_V(z) = \frac{D_V(z)}{r_d},
    \end{align}
    where
    \begin{align}
        D_V(z) = \left[ z\, D_M(z)^2\, D_H(z) \right]^{1/3}, \qquad
        D_M(z) = \int_0^z \frac{c}{H(z')} \, \mathrm{d}z' \, .
    \end{align}

    \item \textbf{Planck CMB} \cite{DESI:2025zgx}: We use compressed likelihoods from Planck 2018 measurements of three key early-Universe parameters: the baryon density $\omega_{\rm b} = \Omega_{\rm b,0} h^2$, the matter density $\omega_m = \Omega_{m,0} h^2$ and the angular size of the sound horizon at recombination $\theta_* = r_* / D_M(z_*)$, 
    where $r_*$ is the comoving sound horizon at photon decoupling and $z_* \approx 1089.8$. We adopt a Gaussian prior on $(\theta_*, \omega_{\rm b}, \omega_m)$ with covariance matrix ${\bm C}$ and mean values $\mu$ from \cite{DESI:2025zgx}:
\begin{align}
    \mu(\theta_{\star}, \omega_{b}, \omega_{m}) = 
    \left(
    \begin{array}{c}
     0.01041  \\
     0.02223 \\
      0.14208  \\
    \end{array}
    \right) \!, ~~ 
    \bm C = 10^{-9} \, 
    \left(
    \begin{array}{ccc}
     0.006621 ~  &  ~ 0.12444 ~ &  ~ -1.1929   \\
     0.12444 ~  &  ~ 21.344 ~  &  ~  -94.001  \\ 
     -1.1929 ~  &  ~ -94.001 ~  &  ~ 1488.4   \\ 
    \end{array}
    \right)\!.
    \end{align}
    
    Note that the comoving sound horizon at redshift $z$ is given by:
\begin{align}
    r_s(z) = \int_z^\infty \frac{c_s(z')}{H(z')} \, \mathrm{d}z' \, ,
\end{align}
where the sound speed in the baryon-photon fluid is given by \cite{Chen:2018dbv}:
\begin{align}
    \frac{c_s(z)^2}{c^2} = \frac{1}{3}\left[1 + \frac{3\rho_{\rm b}(z)}{4\rho_\gamma(z)}\right]^{-1}
    = \frac{1}{3}\left[1 + \frac{3}{4 \Omega_{\gamma, 0} \, h^{2}} \, \frac{\Omega_{\rm b,0}  h^{2}}{1 + z} \right]^{-1} \, .
\end{align}
We compute the sound horizon at the drag epoch, $r_d = r_s(z_d)$, where $z_d \sim 1060$, by integrating numerically. The CMB temperature fixes the photon density today via:
\begin{align}
\frac{3}{4 \Omega_{\gamma, 0} h^2} \approx 31500 \left(\frac{T_{\text{CMB}}}{2.7\,\text{K}}\right)^{-4} \, . 
\end{align}
    
    \item \textbf{Pantheon+ SNIa} \cite{Brout:2022vxf}: We include type Ia supernova data from the Pantheon+ sample, which provides measurements of the apparent peak magnitude $m_B(z)$. These are related to the luminosity distance $D_L(z)$ via:
    \begin{align}
        m_B(z) = M_B + 5 \log_{10} \left[ \frac{D_L(z)}{\text{Mpc}} \right] + 25, \qquad
        D_L(z) = (1 + z) D_M(z),
    \end{align}
    where $M_B$ is the absolute peak magnitude of the type Ia supernovae.
\end{itemize}


\section{Results and discussion}
\label{sec:results}

\begin{figure}[!htbp]
\centering
\includegraphics[width=\linewidth]
{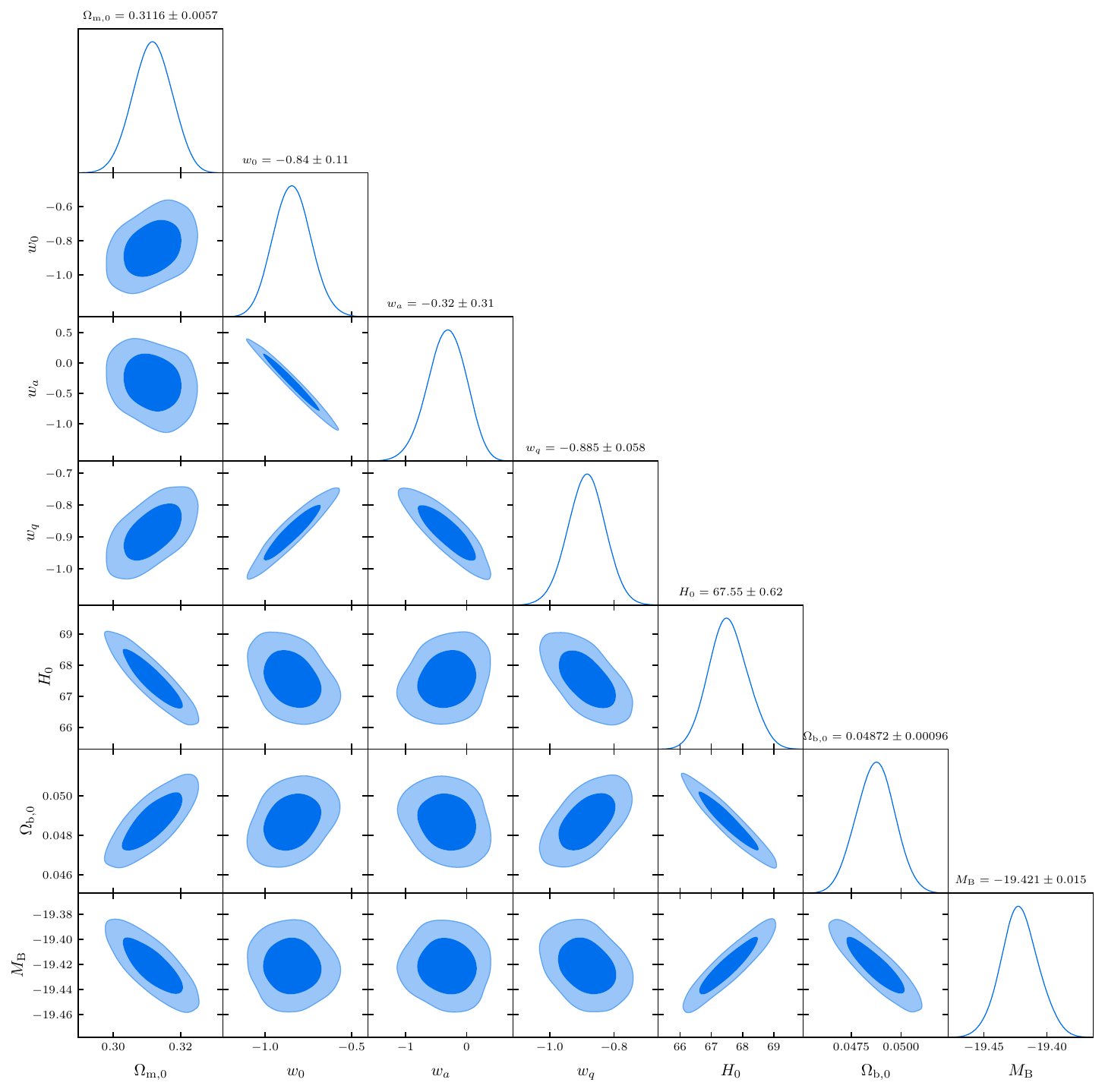} 
\caption{Triangle plot for DESI DR2 BAO + CMB + Pantheon+ constraints. 
\label{fig:triangle}
}
\end{figure}

Using the data combination DESI DR2 BAO  + CMB + Pantheon+, we find the contour plots in \autoref{fig:triangle}, corresponding to the posterior distributions of the parameters. The mean and standard deviation values are given in the plot. These values are used to propagate uncertainties in the quantities of interest {with all possible cross-correlations, which we have not explicitly quoted}.

\begin{figure}[!htbp]
\centering
\includegraphics[width=0.495\linewidth]{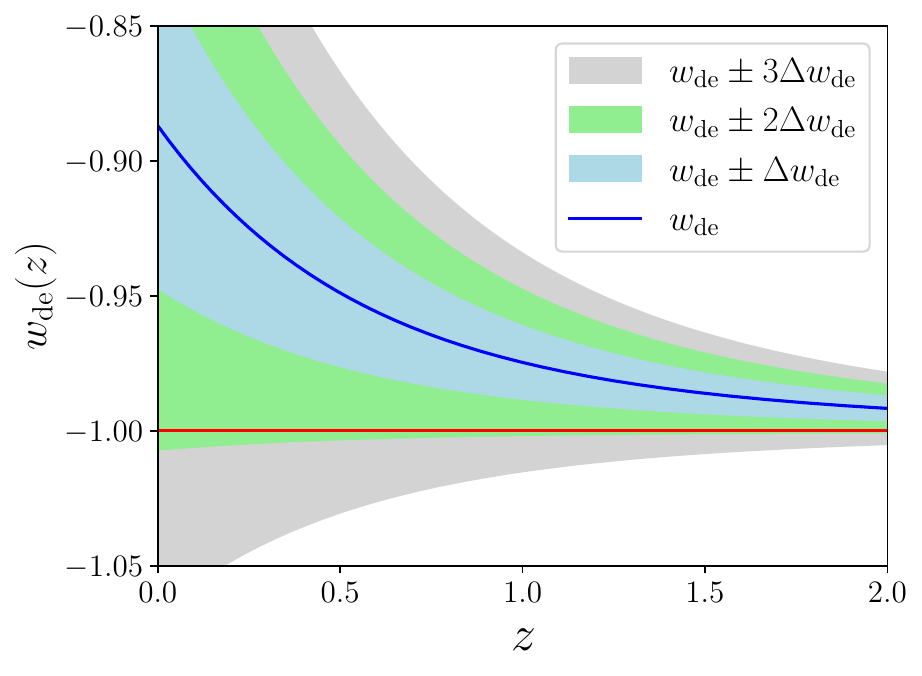}
\includegraphics[width=0.495\linewidth]{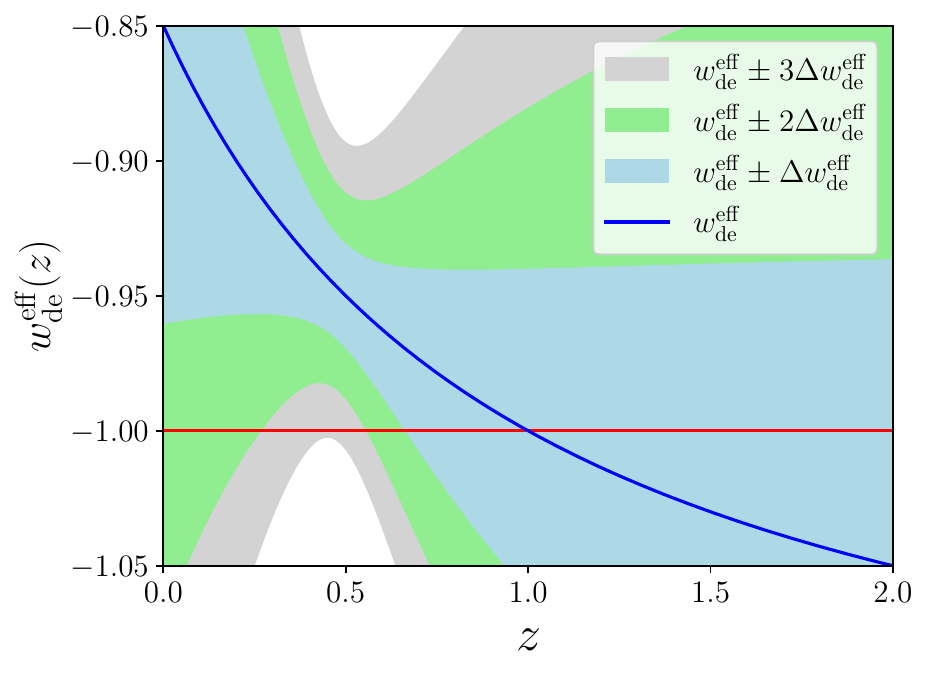}
\caption{Intrinsic ({\em left}) and effective ({\em right}) equations of state for dark energy, with uncertainties. The $\Lambda$ model is the straight red line.
}
\label{fig:wde_wdeff}
\end{figure}

The intrinsic and effective equations of state of the dark energy $w_{\rm de}(z)$ and $w_{\rm de}^{\rm eff}(z)$, are shown in \autoref{fig:wde_wdeff}. Note that for $w_{\rm de}$ we did {\em not} impose a $w_{\rm q}\ge -1$ prior. Although the parametrization \autoref{eq:w_de_thawing} 
does not include this condition (unlike the {actual} quintessence model {with scalar field evolution}), the data shows a preference for non-phantom behaviour of the intrinsic dark energy ({see \cite{Dinda:2024ktd,Dinda:2025iaq} for a similar result in the non-interacting case}). This means that thawing quintessence  with interaction term is a good fit to the current data.

The right panel of \autoref{fig:wde_wdeff} shows how the effective equation of state behaves for an effective non-interacting model of dark energy. As expected from current trends of  cosmological data analysis, the effective behaviour is non-phantom at lower redshifts $(z\lesssim1)$ and phantom at higher redshifts. We note that the non-phantom behaviour at lower redshifts is moderately significant (up to 3$\sigma$), but the phantom behaviour at higher redshifts is not significant (well within 1$\sigma$).

\begin{figure}[!htbp]
\centering
\includegraphics[width=0.495\linewidth]{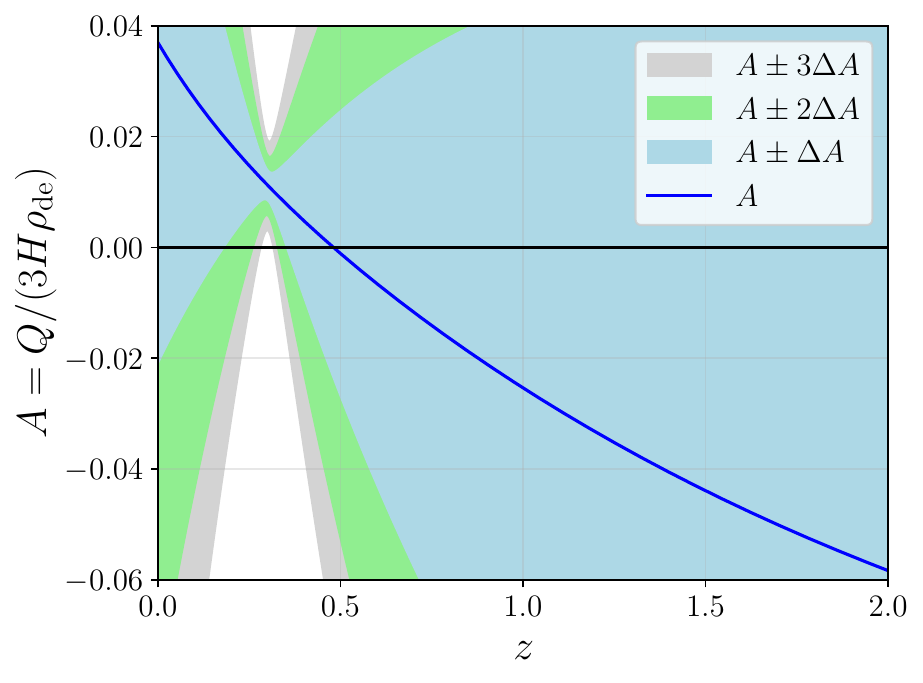}
\caption{Evolution of the interaction, relative to the contribution of intrinsic dark energy.}
\label{fig:A}
\end{figure}

\autoref{fig:A} displays the evolution of the dark interaction relative to the contribution of the intrinsic dark energy, i.e.,  $A=Q/(3H\rho_{\rm de})$.
By \autoref{eq:wdeff}, this is equal to the difference between the effective and intrinsic dark energy equations of state: $A(z)=w^{\rm eff}_{\rm de}(z) - w_{\rm de}(z)$. It is evident that the interaction is negative for higher redshift ($z\gtrsim 0.5$)  and then positive at lower redshift ($z\lesssim 0.5$). In other words, energy is transferred from dark matter to dark energy at early times and then at late times the transfer is reversed. {The negative value of $A$   at higher redshifts ($z\gtrsim 0.5$) is not significant (well within 1$\sigma$), so that $A>0$ for all $z$ is compatible with the data. Notably, $A$ is positive at $z\sim 0.3$, and zero interaction is ruled out at more than $3\sigma$.}

\begin{figure}[!htbp]
\centering
\includegraphics[width=0.51\linewidth]{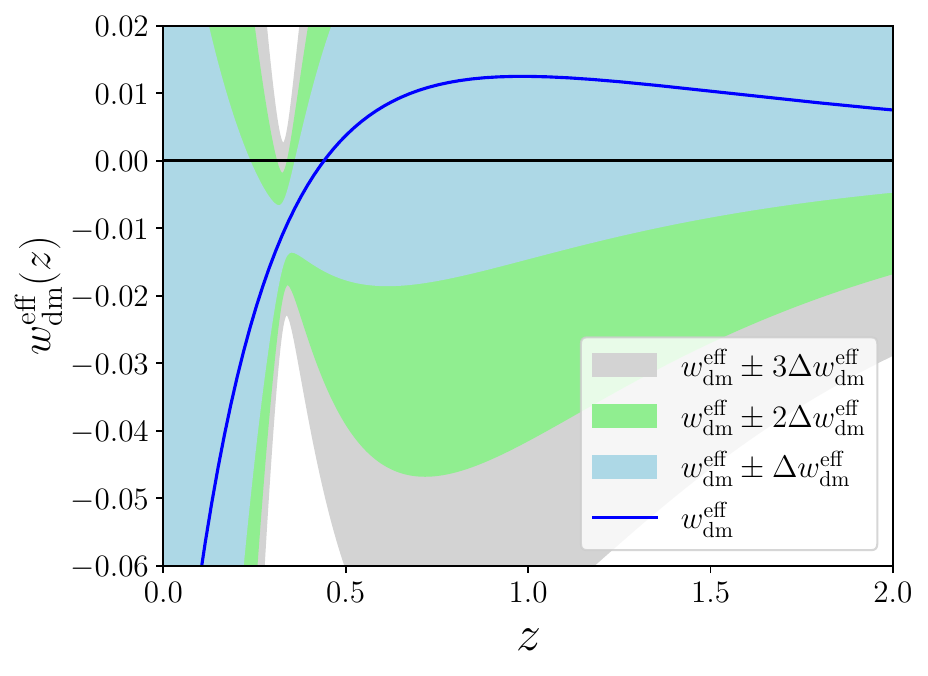}
\includegraphics[width=0.48\linewidth]{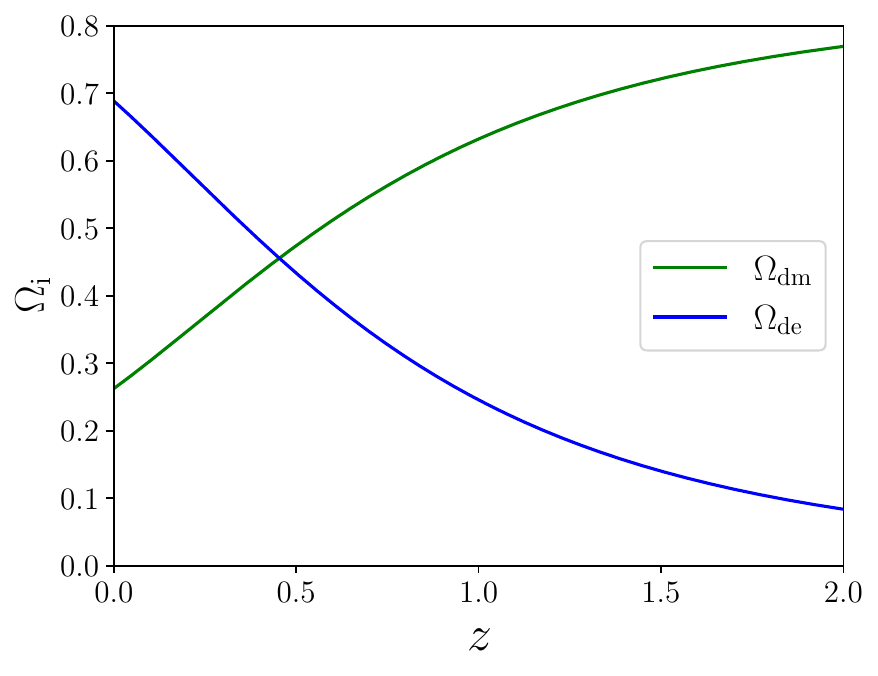}  
\caption{ Evolution of the effective dark matter equation of state ({\em left}) and the dark sector energy densities ({\em right}).
}
\label{fig:wdmeff_OmOq}
\end{figure}

For dark matter, an interaction  can be equivalently described by an effective non-interacting and non-cold dark matter, i.e., with effective nonzero equation of state,  $w_{\rm dm}^{\rm eff}(z)$ \cite{Braglia:2025gdo,Kumar:2025etf}. This is shown in the left panel of \autoref{fig:wdmeff_OmOq}. There is no significant evidence for {nonzero} $w_{\rm dm}^{\rm eff}(z)$,  {except} for  $\sim2\sigma$ evidence for $w_{\rm dm}^{\rm eff}<0$ at $z\sim0.3$.

Finally, in the right panel of \autoref{fig:wdmeff_OmOq}, we compare $\Omega_{\rm dm}$ and $\Omega_{\rm de}$ to see that the dark matter and dark energy equality happens at $z\sim0.5$. All the significant changes happen close to this redshift, as seen in the previous plots. This might be a hint of an intrinsic correlation between dark matter-dark energy equality and the redshift where interaction energy transfer changes sign.

\begin{figure}[!htbp]
\centering
\includegraphics[width=0.49\linewidth]
{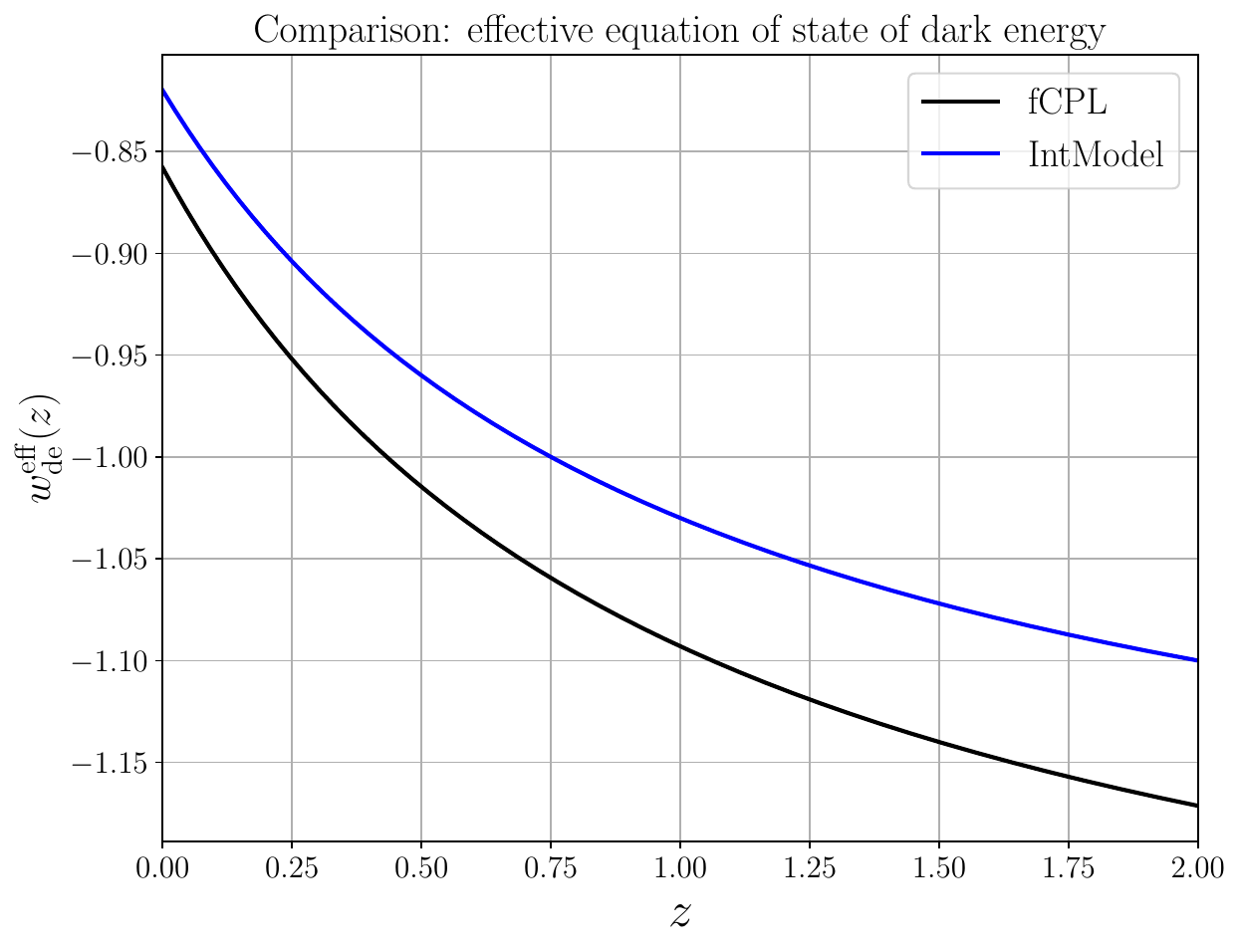}
\includegraphics[width=0.49\linewidth]
{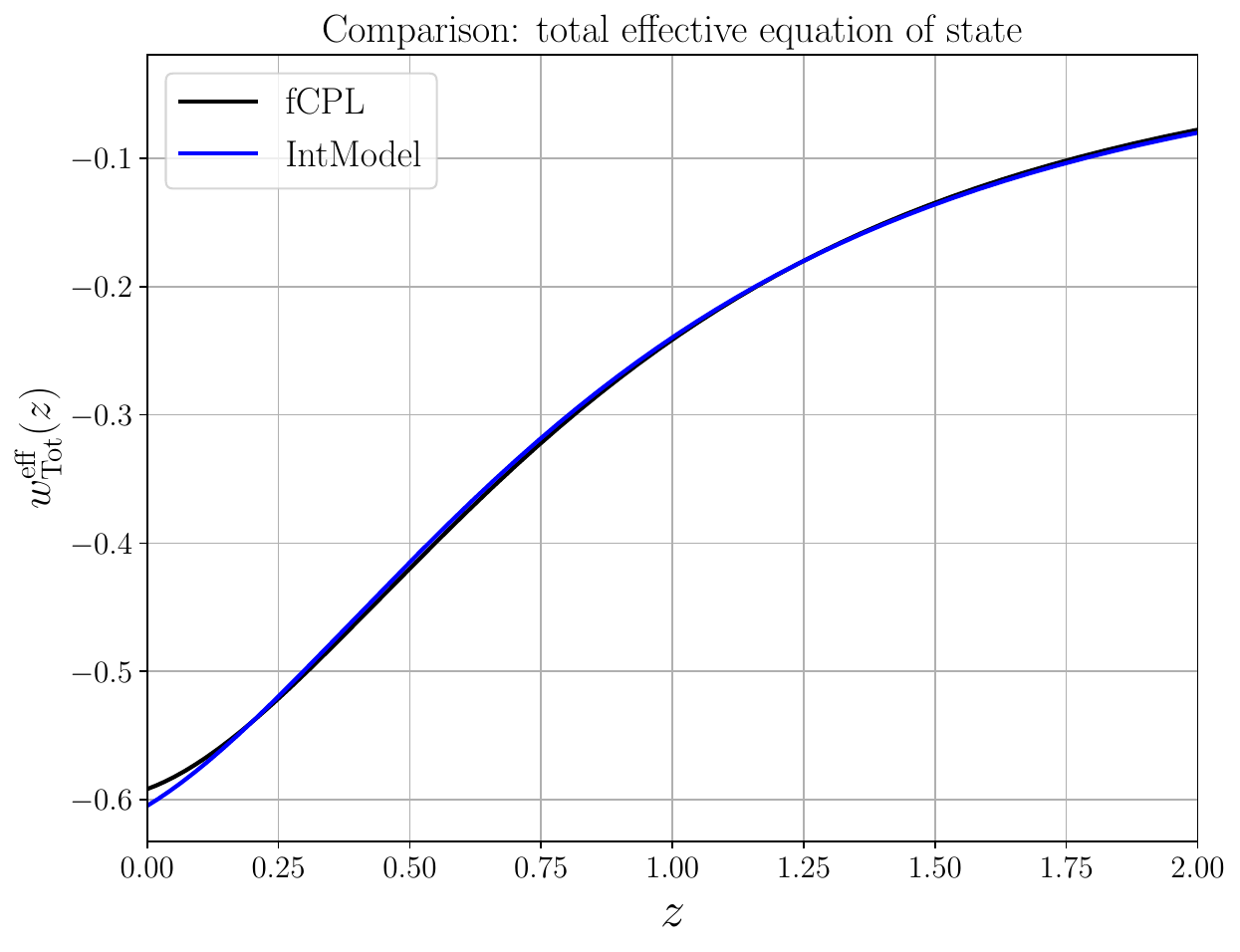}
\caption{
Comparison of the effective equations of state in the interacting model and flat CPL. 
\label{fig:cmp_with_cpl}
}
\end{figure}

\blue{In \autoref{fig:cmp_with_cpl}, we show that in the different models IntModel and fCPL, the effective equation of state of dark energy (left panel) is different but the total effective equation of state (right panel) is  similar. This is because the observed data is more closely related to the {\em total} effective equation of state than the effective equation of state of dark energy.}

\blue{We now make a model comparison between the interacting model considered in this analysis (denoted IntModel), with the standard flat CPL parametrization (fCPL).}

\begin{table}[h!]
\centering
\caption{\blue{Comparison of model selection criteria. Each column lists the difference in the corresponding statistic relative to our interacting model, as in \autoref{eq:model_selection}.}}
\label{table:model_selection}
\vspace{0.2 cm}
\begin{tabular}{|p{2.5cm}|p{3.5cm}|p{2.5cm}|p{2.5cm}|p{2.5cm}|}
\hline
\qquad $\Delta\chi^2$ & \qquad $\Delta \ln \mathcal{Z}$ & \qquad $\Delta$AIC & \qquad $\Delta$BIC & \qquad $\Delta$DIC \\
\hline
\qquad 1.539 & \qquad -4.934 & \qquad 3.536 & \qquad 8.918 & \qquad -0.944 \\
\hline
\end{tabular}
\end{table}

\noindent \blue{In \autoref{table:model_selection}, we show the relative difference for a given statistic $S$, defined as}
\begin{equation}
\blue{
\Delta S = S(\text{IntModel}) - S(\text{fCPL}) \, .
}
\label{eq:model_selection}
\end{equation}
\blue{where \cite{Trotta:2008qt,Liddle:2007fy}}
\begin{eqnarray}
\Delta \ln \mathcal{Z} &=& \ln \mathcal{Z}_{\text{IntModel}} - \ln \mathcal{Z}_{\text{fCPL}} \, ,
\label{eq:BayesFactor} \\
\mathrm{AIC} &=& -2\ln\mathcal{L}_{\mathrm{max}} + 2k \, ,
\label{eq:AIC} \\
\mathrm{BIC} &=& -2\ln\mathcal{L}_{\mathrm{max}} + k\ln N \, ,
\label{eq:BIC} \\
\mathrm{DIC} &=& -2\ln\mathcal{L}(\bar{\theta}) + 2p_D \quad \mbox{where}~~ p_D = -2 ~ \overline{\ln\mathcal{L}(\theta)} + 2\ln\mathcal{L}(\bar{\theta}) \, .
\label{eq:DIC}
\end{eqnarray}
\blue{The quantities  above are defined as follows (see \autoref{sec:model_selection} for further details).
\begin{itemize}
    \item $\mathcal{Z}$ is the Bayesian evidence, i.e., the marginal likelihood of the data given the model.
    \item $\mathcal{L}_{\mathrm{max}}$ is the maximum likelihood over the parameter space.
    \item $k$ is the total number of free parameters in the model.
    \item $N$ is the number of data points used in the fit.
    \item $\bar{\theta}$ is the posterior mean of the model parameters.
    \item $p_D$ is the effective number of parameters, quantifying model complexity.
    \item $\mathcal{L}(\theta)$ is the likelihood evaluated at a given parameter set $\theta$.
    \item $\overline{\ln\mathcal{L}(\theta)}$ is the posterior mean of the log-likelihood, which can be approximated as
\begin{equation}
\overline{\ln\mathcal{L}(\theta)} \approx \frac{1}{N_{\mathrm{MC}}} \sum_{i=1}^{N_{\mathrm{MC}}} \ln\mathcal{L}(\theta_i)\,,
\label{eq:posterior_avg_loglike}
\end{equation}
where $\{\theta_i\}$ are the posterior samples and $N_{\mathrm{MC}}$ is the total number of samples.
\end{itemize}
\noindent
The  preference of our IntModel compared to the fCPL model based on the different selection criteria is summarized as follows.
\begin{itemize}
    \item fCPL  provides a slightly better fit to the data according to $\chi^2$.
    \item Bayesian evidence shows moderate preference for the fCPL.
    \item AIC favours fCPL, though the difference is not statistically significant.
    \item BIC strongly prefers fCPL.
    \item DIC marginally favors  IntModel, but the difference is negligible.
\end{itemize}
Except for the BIC criterion, the IntModel has equivalent evidence to fCPL. The BIC favours fCPL because it strongly penalizes the IntModel, which has an extra parameter, $w_q$. In summary, the data does not show significant evidence for a preference of fCPL over IntModel. 
}


\section{Conclusion} 
\label{sec:conclusion}

The $w_0,w_a$ parametrization of the dark energy equation state leads to claims of phantom crossing on the basis of the latest cosmological data. We tested the idea that phantom behaviour -- which is physically inconsistent in general relativity -- is not real, but only apparent. In other words, it could be a signal of interaction in the dark sector. For this purpose, we assumed  a physically self-consistent model of {intrinsic} dark energy, in the form of thawing quintessence-like parametrization, and we assumed a $w_0,w_a$ parametrization of the effective equation of state of dark energy. Then we  reconstructed the redshift dependence of the interaction function, $Q(z)$.

The parametrization of the {intrinsic} equation of state of dark energy behaves like thawing quintessence for $w_{\rm q}>-1$\blue{, where $w_{\rm q}$ is the present value of the intrinsic equation of state of dark energy.} On the other hand, it has phantom behavior for $w_{\rm q}<-1$. We did not impose a $w_{\rm q}>-1$ prior, but we still found non-phantom values of $w_{\rm de}(z)$ from current data, hinting that thawing quintessence models may be a good fit to the data in dark interaction models. The effective equation of state of dark energy suggests that there is no significant evidence for phantom crossing even in $w_{\rm de}^{\rm eff}$, which is assumed to be a $w_0,w_a$ parametrization.

We find that the interaction function $Q(z)$ shows a preference for negative values (although not significant) at higher redshifts ($z\gtrsim0.5$) and changes sign around $z\sim0.5$ to become positive at lower redshifts. This suggests an energy flow from dark matter to dark energy in the early Universe and energy flow from dark energy to dark matter in the late Universe. Interestingly, this transition happens close to the era of dark matter-dark energy equality.

We also find that at $z\sim0.3$ the effective equation of state of dark matter $w_{\rm dm}^{\rm eff}$ is negative at around 2$\sigma$. At other redshifts, there is no significant evidence for nonzero $w_{\rm dm}^{\rm eff}$.

In summary, we find interacting dark energy models can be viable alternatives to both the standard $\Lambda$CDM model and phantom dark energy scenarios, while the intrinsic dark energy equation of state is non-phantom.


\acknowledgments
The authors are supported by the South African Radio Astronomy Observatory and the National Research Foundation (grant no. 75415). 


\appendix

\section{Model selection criteria}
\label{sec:model_selection}

\blue{Here we briefly discuss the model selection criteria and decision rules associated with the statistical quantities $\chi^2$ and those defined in \autoref{eq:BayesFactor}--\autoref{eq:DIC}. We use the convention}
\begin{equation}
\Delta S = S(\text{Model 1}) - S(\text{Model 2})\,,
\label{eq:S_difference}
\end{equation}
\blue{so that positive (negative) values of $\Delta S$ mean the statistic is larger (smaller) for Model 1 than for Model 2. For each model comparison criterion, the tables below present the conventional scale for the strength of evidence for one model over another.

The decision rules for different criteria are as follows.
\begin{itemize}
    \item  $\Delta\chi^2 < 0$,  $\Delta\mathrm{AIC}<0$, $\Delta\mathrm{BIC}<0$, $\Delta\mathrm{DIC}<0$ all indicate stronger evidence in favour of  Model 1  \cite{Pearson1992,Robert_2009,KassRaftery1995,1100705,Schwarz1978}.
    \item  $\Delta \ln \mathcal{Z} > 0$ favours Model~1
\cite{Robert_2009,KassRaftery1995,Spiegelhalter2002}.
\end{itemize}
The quantitative criterion scale for model selection is listed in the tables below for each criterion.
}

\begin{table}[h!]
\centering
\caption{Interpretation of $\Delta \ln \mathcal{Z}$ (Bayes evidence difference) according to the Jeffreys scale.}
\label{table:bayes_evidence_scale}
\vspace{0.2 cm}
\begin{tabular}{|p{3cm}|p{10cm}|}
\hline
\qquad $\Delta \ln \mathcal{Z}$ &  Strength of evidence (Model 1 vs Model 2) \\
\hline
\qquad $0$ -- $1$ &  No substantial evidence for favouring Model 1 or 2 \\
\qquad $1$ -- $2.5$ &  Weak evidence favouring Model 1 over 2 \\
\qquad $2.5$ -- $5$ &  Moderate evidence favouring Model 1 over  2 \\
\qquad $> 5$ &  Strong evidence favouring Model 1 over  2 \\
\hline
\end{tabular}
\end{table}

\begin{table}[h!]
\centering
\caption{Interpretation of $\Delta\mathrm{AIC} = \mathrm{AIC}(\text{Model 1}) - \mathrm{AIC}(\text{Model 2})$.}
\label{table:aic_scale}
\vspace{0.2 cm}
\begin{tabular}{|p{3cm}|p{10cm}|}
\hline
\qquad $\Delta \mathrm{AIC}$ &  Strength of evidence (Model 1 vs Model 2) \\
\hline
\qquad $0$ -- $2$ & Model 1 slightly disfavoured relative to  2 (negligible to weak) \\
\qquad $4$ -- $7$ &  Model 1 considerably disfavoured relative to  2 (moderate evidence against Model 1) \\
\qquad $> 10$ &  Model 1 strongly disfavoured relative to  2 (essentially no support for Model 1) \\
\hline
\end{tabular}
\end{table}

\begin{table}[h!]
\centering
\caption{Interpretation of $\Delta\mathrm{BIC} = \mathrm{BIC}(\text{Model 1}) - \mathrm{BIC}(\text{Model 2})$.}
\label{table:bic_scale}
\vspace{0.2 cm}
\begin{tabular}{|p{3cm}|p{10cm}|}
\hline
\qquad $\Delta \mathrm{BIC}$ &  Strength of evidence (Model 1 vs Model 2) \\
\hline
\qquad $0$ -- $2$ & Model 1 weakly disfavoured relative to  2 \\
\qquad $2$ -- $6$ & Model 1 moderately disfavoured relative to  2 \\
\qquad $6$ -- $10$ &  Model 1 strongly disfavoured relative to  2 \\
\qquad $> 10$ &  Model 1 very strongly disfavoured relative to  2 \\
\hline
\end{tabular}
\end{table}

\begin{table}[h!]
\centering
\caption{Interpretation of $\Delta\mathrm{DIC} = \mathrm{DIC}(\text{Model 1}) - \mathrm{DIC}(\text{Model 2})$.}
\label{table:dic_scale}
\vspace{0.2 cm}
\begin{tabular}{|p{3cm}|p{10cm}|}
\hline
\qquad $\Delta \mathrm{DIC}$ &  Strength of evidence (Model 1 vs Model 2) \\
\hline
\qquad $0$ -- $2$ &  
Model 1 slightly disfavoured relative to  2 \\
\qquad $2$ -- $7$ & Model 1 moderately disfavoured relative to  2 \\
\qquad $> 10$ &  Model 1 strongly disfavoured relative to  2 \\
\hline
\end{tabular}
\end{table}

\clearpage
\bibliographystyle{JHEP}
\bibliography{reference_library}

@article{Yang:2015tzc,
    author = "Yang, Tao and Guo, Zong-Kuan and Cai, Rong-Gen",
    title = "{Reconstructing the interaction between dark energy and dark matter using Gaussian Processes}",
    eprint = "1505.04443",
    archivePrefix = "arXiv",
    primaryClass = "astro-ph.CO",
    doi = "10.1103/PhysRevD.91.123533",
    journal = "Phys. Rev. D",
    volume = "91",
    number = "12",
    pages = "123533",
    year = "2015"
}

@article{Zimdahl:2012mj,
    author = "Zimdahl, W.",
    editor = "Alcaniz, Jailson and Carneiro, Saulo and Chimento, Luis P. and Del Campo, Sergio and Fabris, Julio C. and Lima, J. A. S. and Zimdahl, Winfried",
    title = "{Models of Interacting Dark Energy}",
    eprint = "1204.5892",
    archivePrefix = "arXiv",
    primaryClass = "astro-ph.CO",
    doi = "10.1063/1.4756811",
    journal = "AIP Conf. Proc.",
    volume = "1471",
    pages = "51--56",
    year = "2012"
}

@article{Pourtsidou:2013nha,
    author = "Pourtsidou, A. and Skordis, C. and Copeland, E. J.",
    title = "{Models of dark matter coupled to dark energy}",
    eprint = "1307.0458",
    archivePrefix = "arXiv",
    primaryClass = "astro-ph.CO",
    doi = "10.1103/PhysRevD.88.083505",
    journal = "Phys. Rev. D",
    volume = "88",
    number = "8",
    pages = "083505",
    year = "2013"
}

@article{Wang:2016lxa,
    author = "Wang, B. and Abdalla, E. and Atrio-Barandela, F. and Pavon, D.",
    title = "{Dark Matter and Dark Energy Interactions: Theoretical Challenges, Cosmological Implications and Observational Signatures}",
    eprint = "1603.08299",
    archivePrefix = "arXiv",
    primaryClass = "astro-ph.CO",
    doi = "10.1088/0034-4885/79/9/096901",
    journal = "Rept. Prog. Phys.",
    volume = "79",
    number = "9",
    pages = "096901",
    year = "2016"
}

@article{DESI:2025fii,
    author = "Lodha, K. and others",
    collaboration = "DESI",
    title = "{Extended Dark Energy analysis using DESI DR2 BAO measurements}",
    eprint = "2503.14743",
    archivePrefix = "arXiv",
    primaryClass = "astro-ph.CO",
    reportNumber = "FERMILAB-PUB-25-0164-PPD",
    month = "3",
    year = "2025"
}

@article{DESI:2025zgx,
    author = "Abdul Karim, M. and others",
    collaboration = "DESI",
    title = "{DESI DR2 Results II: Measurements of Baryon Acoustic Oscillations and Cosmological Constraints}",
    eprint = "2503.14738",
    archivePrefix = "arXiv",
    primaryClass = "astro-ph.CO",
    reportNumber = "FERMILAB-PUB-25-0169-PPD",
    month = "3",
    year = "2025"
}

@article{Yang:2025boq,
    author = "Yang, Yupeng and Dai, Xinyi and Wang, Yicheng",
    title = "{New cosmological constraints on the evolution of dark matter energy density}",
    eprint = "2505.09879",
    archivePrefix = "arXiv",
    primaryClass = "astro-ph.CO",
    month = "5",
    year = "2025"
}

@article{Zhai:2025hfi,
    author = "Zhai, Yuejia and de Cesare, Marco and van de Bruck, Carsten and Di Valentino, Eleonora and Wilson-Ewing, Edward",
    title = "{A low-redshift preference for an interacting dark energy model}",
    eprint = "2503.15659",
    archivePrefix = "arXiv",
    primaryClass = "astro-ph.CO",
    month = "3",
    year = "2025"
}

@article{Dinda:2024ktd,
    author = "Dinda, Bikash R. and Maartens, Roy",
    title = "{Model-agnostic assessment of dark energy after DESI DR1 BAO}",
    eprint = "2407.17252",
    archivePrefix = "arXiv",
    primaryClass = "astro-ph.CO",
    doi = "10.1088/1475-7516/2025/01/120",
    journal = "JCAP",
    volume = "01",
    pages = "120",
    year = "2025"
}

@article{Dinda:2025iaq,
    author = "Dinda, Bikash R. and Maartens, Roy",
    title = "{Physical vs phantom dark energy after DESI: thawing quintessence in a curved background}",
    eprint = "2504.15190",
    archivePrefix = "arXiv",
    primaryClass = "astro-ph.CO",
    doi = "10.1093/mnrasl/slaf063",
    journal = "Mon. Not. Roy. Astron. Soc. Lett.",
    volume = "542",
    pages = "L31",
    year = "2025"
}

@article{Escamilla:2023shf,
    author = "Escamilla, Luis A. and Akarsu, Ozgur and Di Valentino, Eleonora and Vazquez, J. Alberto",
    title = "{Model-independent reconstruction of the interacting dark energy kernel: Binned and Gaussian process}",
    eprint = "2305.16290",
    archivePrefix = "arXiv",
    primaryClass = "astro-ph.CO",
    doi = "10.1088/1475-7516/2023/11/051",
    journal = "JCAP",
    volume = "11",
    pages = "051",
    year = "2023"
}

@article{Abedin:2025yru,
    author = "Abedin, Mazaharul and Wang, Guo-Jian and Ma, Yin-Zhe and Pan, Supriya",
    title = "{In search of an interaction in the dark sector through Gaussian Process and ANN approaches}",
    eprint = "2505.04336",
    archivePrefix = "arXiv",
    primaryClass = "astro-ph.CO",
    month = "5",
    year = "2025"
}

@article{Yashiki:2025loj,
    author = "Yashiki, Mai",
    title = "{Toward a simultaneous resolution of the $H_0$ and $S_8$ tensions: early dark energy and an interacting dark sector model}",
    eprint = "2505.23382",
    archivePrefix = "arXiv",
    primaryClass = "astro-ph.CO",
    month = "5",
    year = "2025"
}

@article{Linder:2025zxb,
    author = "Linder, Eric V.",
    title = "{Uplifting, Depressing, and Tilting Dark Energy}",
    eprint = "2506.02122",
    archivePrefix = "arXiv",
    primaryClass = "astro-ph.CO",
    month = "6",
    year = "2025"
}

@article{Li:2025ula,
    author = "Li, Yun-He and Zhang, Xin",
    title = "{Cosmic Sign-Reversal: Non-Parametric Reconstruction of Interacting Dark Energy with DESI DR2}",
    eprint = "2506.18477",
    archivePrefix = "arXiv",
    primaryClass = "astro-ph.CO",
    month = "6",
    year = "2025"
}

@article{Chen:2018dbv,
    author = "Chen, Lu and Huang, Qing-Guo and Wang, Ke",
    title = "{Distance Priors from Planck Final Release}",
    eprint = "1808.05724",
    archivePrefix = "arXiv",
    primaryClass = "astro-ph.CO",
    doi = "10.1088/1475-7516/2019/02/028",
    journal = "JCAP",
    volume = "02",
    pages = "028",
    year = "2019"
}

@article{Brout:2022vxf,
    author = "Brout, Dillon and others",
    title = "{The Pantheon+ Analysis: Cosmological Constraints}",
    eprint = "2202.04077",
    archivePrefix = "arXiv",
    primaryClass = "astro-ph.CO",
    doi = "10.3847/1538-4357/ac8e04",
    journal = "Astrophys. J.",
    volume = "938",
    number = "2",
    pages = "110",
    year = "2022"
}

@article{delaMacorra:2021hoh,
    author = "de la Macorra, Axel and Almaraz, Erick and Garrido, Joanna",
    title = "{Towards a solution to the H0 tension}",
    eprint = "2106.12116",
    archivePrefix = "arXiv",
    primaryClass = "astro-ph.CO",
    doi = "10.1103/PhysRevD.105.023526",
    journal = "Phys. Rev. D",
    volume = "105",
    number = "2",
    pages = "023526",
    year = "2022"
}

@article{DiValentino:2021izs,
    author = "Di Valentino, Eleonora and Mena, Olga and Pan, Supriya and Visinelli, Luca and Yang, Weiqiang and Melchiorri, Alessandro and Mota, David F. and Riess, Adam G. and Silk, Joseph",
    title = "{In the realm of the Hubble tension{\textemdash}a review of solutions}",
    eprint = "2103.01183",
    archivePrefix = "arXiv",
    primaryClass = "astro-ph.CO",
    reportNumber = "IPPP/20/108",
    doi = "10.1088/1361-6382/ac086d",
    journal = "Class. Quant. Grav.",
    volume = "38",
    number = "15",
    pages = "153001",
    year = "2021"
}

@article{Verde:2019ivm,
    author = "Verde, L. and Treu, T. and Riess, A. G.",
    title = "{Tensions between the Early and the Late Universe}",
    eprint = "1907.10625",
    archivePrefix = "arXiv",
    primaryClass = "astro-ph.CO",
    doi = "10.1038/s41550-019-0902-0",
    journal = "Nature Astron.",
    volume = "3",
    pages = "891",
    year = "2019"
}

@article{DES:2021wwk,
    author = "Abbott, T. M. C. and others",
    collaboration = "DES",
    title = "{Dark Energy Survey Year 3 results: Cosmological constraints from galaxy clustering and weak lensing}",
    eprint = "2105.13549",
    archivePrefix = "arXiv",
    primaryClass = "astro-ph.CO",
    reportNumber = "FERMILAB-PUB-21-221-AE, DES-2020-0617",
    doi = "10.1103/PhysRevD.105.023520",
    journal = "Phys. Rev. D",
    volume = "105",
    number = "2",
    pages = "023520",
    year = "2022"
}

@article{Hildebrandt:2016iqg,
    author = "Hildebrandt, H. and others",
    title = "{KiDS-450: Cosmological parameter constraints from tomographic weak gravitational lensing}",
    eprint = "1606.05338",
    archivePrefix = "arXiv",
    primaryClass = "astro-ph.CO",
    doi = "10.1093/mnras/stw2805",
    journal = "Mon. Not. Roy. Astron. Soc.",
    volume = "465",
    pages = "1454",
    year = "2017"
}

@article{Zlatev:1998tr,
    author = "Zlatev, Ivaylo and Wang, Li-Min and Steinhardt, Paul J.",
    title = "{Quintessence, cosmic coincidence, and the cosmological constant}",
    eprint = "astro-ph/9807002",
    archivePrefix = "arXiv",
    doi = "10.1103/PhysRevLett.82.896",
    journal = "Phys. Rev. Lett.",
    volume = "82",
    pages = "896--899",
    year = "1999"
}

@inproceedings{Weinberg:2000yb,
    author = "Weinberg, Steven",
    title = "{The Cosmological constant problems}",
    booktitle = "{4th International Symposium on Sources and Detection of Dark Matter in the Universe (DM 2000)}",
    eprint = "astro-ph/0005265",
    archivePrefix = "arXiv",
    reportNumber = "UTTG-07-00",
    doi = "10.1007/978-3-662-04587-9_2",
    pages = "18--26",
    month = "2",
    year = "2000"
}

@article{Planck:2018vyg,
    author = "Aghanim, N. and others",
    collaboration = "Planck",
    title = "{Planck 2018 results. VI. Cosmological parameters}",
    eprint = "1807.06209",
    archivePrefix = "arXiv",
    primaryClass = "astro-ph.CO",
    doi = "10.1051/0004-6361/201833910",
    journal = "Astron. Astrophys.",
    volume = "641",
    pages = "A6",
    year = "2020",
    note = "[Erratum: Astron.Astrophys. 652, C4 (2021)]"
}

@article{Bertone:2004pz,
    author = "Bertone, Gianfranco and Hooper, Dan and Silk, Joseph",
    title = "{Particle dark matter: Evidence, candidates and constraints}",
    eprint = "hep-ph/0404175",
    archivePrefix = "arXiv",
    reportNumber = "FERMILAB-PUB-04-047-A",
    doi = "10.1016/j.physrep.2004.08.031",
    journal = "Phys. Rept.",
    volume = "405",
    pages = "279--390",
    year = "2005"
}

@article{Frieman:2008sn,
    author = "Frieman, Joshua and Turner, Michael and Huterer, Dragan",
    title = "{Dark Energy and the Accelerating Universe}",
    eprint = "0803.0982",
    archivePrefix = "arXiv",
    primaryClass = "astro-ph",
    reportNumber = "FERMILAB-PUB-08-613-A",
    doi = "10.1146/annurev.astro.46.060407.145243",
    journal = "Ann. Rev. Astron. Astrophys.",
    volume = "46",
    pages = "385--432",
    year = "2008"
}

@article{Chakraborty:2025syu,
    author = "Chakraborty, Amlan and Chanda, Prolay K. and Das, Subinoy and Dutta, Koushik",
    title = "{DESI results: Hint towards coupled dark matter and dark energy}",
    eprint = "2503.10806",
    archivePrefix = "arXiv",
    primaryClass = "astro-ph.CO",
    month = "3",
    year = "2025"
}

@article{Shah:2025ayl,
    author = "Shah, Rahul and Mukherjee, Purba and Pal, Supratik",
    title = "{Interacting dark sectors in light of DESI DR2}",
    eprint = "2503.21652",
    archivePrefix = "arXiv",
    primaryClass = "astro-ph.CO",
    month = "3",
    year = "2025"
}

@article{Braglia:2025gdo,
    author = "Braglia, Matteo and Chen, Xingang and Loeb, Abraham",
    title = "{Exotic Dark Matter and the DESI Anomaly}",
    eprint = "2507.13925",
    archivePrefix = "arXiv",
    primaryClass = "astro-ph.CO",
    month = "7",
    year = "2025"
}

@article{Montani:2025rcy,
    author = "Montani, Giovanni and Fazzari, Elisa and Carlevaro, Nakia and Dainotti, Maria Giovanna",
    title = "{Two dynamical scenarios for the binned Master sample interpretation}",
    eprint = "2507.14048",
    archivePrefix = "arXiv",
    primaryClass = "astro-ph.CO",
    month = "7",
    year = "2025"
}

@article{Dinda:2022ixi,
    author = "Dinda, Bikash R. and Hossain, Md. Wali and Sen, Anjan A.",
    title = "{21-cm power spectrum in interacting cubic Galileon model}",
    eprint = "2208.11560",
    archivePrefix = "arXiv",
    primaryClass = "astro-ph.CO",
    doi = "10.1007/s12036-023-09976-2",
    journal = "J. Astrophys. Astron.",
    volume = "44",
    number = "2",
    pages = "85",
    year = "2023"
}

@article{Valiviita:2009nu,
    author = "Valiviita, Jussi and Maartens, Roy and Majerotto, Elisabetta",
    title = "{Observational constraints on an interacting dark energy model}",
    eprint = "0907.4987",
    archivePrefix = "arXiv",
    primaryClass = "astro-ph.CO",
    doi = "10.1111/j.1365-2966.2009.16115.x",
    journal = "Mon. Not. Roy. Astron. Soc.",
    volume = "402",
    pages = "2355--2368",
    year = "2010"
}

@article{Clemson:2011an,
    author = "Clemson, Timothy and Koyama, Kazuya and Zhao, Gong-Bo and Maartens, Roy and Valiviita, Jussi",
    title = "{Interacting Dark Energy -- constraints and degeneracies}",
    eprint = "1109.6234",
    archivePrefix = "arXiv",
    primaryClass = "astro-ph.CO",
    doi = "10.1103/PhysRevD.85.043007",
    journal = "Phys. Rev. D",
    volume = "85",
    pages = "043007",
    year = "2012"
}

@article{Dinda:2021ffa,
    author = "Dinda, Bikash R.",
    title = "{Cosmic expansion parametrization: Implication for curvature and H0 tension}",
    eprint = "2106.02963",
    archivePrefix = "arXiv",
    primaryClass = "astro-ph.CO",
    reportNumber = "TIFR/TH/21-18",
    doi = "10.1103/PhysRevD.105.063524",
    journal = "Phys. Rev. D",
    volume = "105",
    number = "6",
    pages = "063524",
    year = "2022"
}

@article{aghanim2020planck,
  title={Planck 2018 results-VI. Cosmological parameters},
  author={Aghanim, Nabila and Akrami, Yashar and Ashdown, Mark and Aumont, Jonathan and Baccigalupi, Carlo and Ballardini, Mario and Banday, Anthony J and Barreiro, RB and Bartolo, Nicola and Basak, S and others},
 journal={Astronomy \& Astrophysics},
  volume={641},
  pages={A6},
  year={2020},
  publisher={EDP sciences}
}

@article{Yan:2025iga,
    author = "Yan, Hang and Pan, Yu and Wang, Jia-Xin and Xu, Wen-Xiao and Peng, Ze-Hui",
    title = "{Investigating Interacting Dark Energy Models Using Fast Radio Burst Observations}",
    eprint = "2507.16308",
    archivePrefix = "arXiv",
    primaryClass = "astro-ph.CO",
    month = "7",
    year = "2025"
}

@article{Brax:2025ahm,
    author = "Brax, Philippe",
    title = "{Weinberg's theorem, phantom crossing and screening}",
    eprint = "2507.16723",
    archivePrefix = "arXiv",
    primaryClass = "astro-ph.CO",
    month = "7",
    year = "2025"
}

@article{Kumar:2025etf,
    author = "Kumar, Utkarsh and Ajith, Abhijith and Verma, Amresh",
    title = "{Evidence for non-cold dark matter from DESI DR2 measurements}",
    eprint = "2504.14419",
    archivePrefix = "arXiv",
    primaryClass = "astro-ph.CO",
    month = "4",
    year = "2025"
}

@article{Li:2024qso,
    author = "Li, Tian-Nuo and Wu, Peng-Ju and Du, Guo-Hong and Jin, Shang-Jie and Li, Hai-Li and Zhang, Jing-Fei and Zhang, Xin",
    title = "{Constraints on Interacting Dark Energy Models from the DESI Baryon Acoustic Oscillation and DES Supernovae Data}",
    eprint = "2407.14934",
    archivePrefix = "arXiv",
    primaryClass = "astro-ph.CO",
    doi = "10.3847/1538-4357/ad87f0",
    journal = "Astrophys. J.",
    volume = "976",
    number = "1",
    pages = "1",
    year = "2024"
}

@article{Li:2025owk,
    author = "Li, Tian-Nuo and Du, Guo-Hong and Li, Yun-He and Wu, Peng-Ju and Jin, Shang-Jie and Zhang, Jing-Fei and Zhang, Xin",
    title = "{Probing the sign-changeable interaction between dark energy and dark matter with DESI baryon acoustic oscillations and DES supernovae data}",
    eprint = "2501.07361",
    archivePrefix = "arXiv",
    primaryClass = "astro-ph.CO",
    month = "1",
    year = "2025"
}

@article{Guin:2025xki,
    author = "Guin, Gopinath and Paul, Souvik and Gangopadhyay, Sunandan",
    title = "{Interacting Barrow holographic dark energy model in the presence of radiation and matter}",
    eprint = "2507.01070",
    archivePrefix = "arXiv",
    primaryClass = "gr-qc",
    month = "7",
    year = "2025"
}

@article{Gialamas:2025pwv,
    author = {Gialamas, Ioannis D. and H{\"u}tsi, Gert and Raidal, Martti and Urrutia, Juan and Vasar, Martin and Veerm{\"a}e, Hardi},
    title = "{Quintessence and phantoms in light of DESI 2025}",
    eprint = "2506.21542",
    archivePrefix = "arXiv",
    primaryClass = "astro-ph.CO",
    month = "6",
    year = "2025"
}

@article{Nojiri:2025low,
    author = "Nojiri, Shin'ichi and Odintsov, S. D. and Oikonomou, V. K.",
    title = "{Phantom Crossing and Oscillating Dark Energy with $F(R)$ Gravity}",
    eprint = "2506.21010",
    archivePrefix = "arXiv",
    primaryClass = "gr-qc",
    reportNumber = "KEK-TH-2734, KEK-Cosmo-0383",
    month = "6",
    year = "2025"
}

@article{RoyChoudhury:2024wri,
    author = "Roy Choudhury, Shouvik and Okumura, Teppei",
    title = "{Updated Cosmological Constraints in Extended Parameter Space with Planck PR4, DESI Baryon Acoustic Oscillations, and Supernovae: Dynamical Dark Energy, Neutrino Masses, Lensing Anomaly, and the Hubble Tension}",
    eprint = "2409.13022",
    archivePrefix = "arXiv",
    primaryClass = "astro-ph.CO",
    doi = "10.3847/2041-8213/ad8c26",
    journal = "Astrophys. J. Lett.",
    volume = "976",
    number = "1",
    pages = "L11",
    year = "2024"
}

@article{RoyChoudhury:2025dhe,
    author = "Roy Choudhury, Shouvik",
    title = "{Cosmology in Extended Parameter Space with DESI Data Release 2 Baryon Acoustic Oscillations: A 2{\ensuremath{\sigma}}+ Detection of Nonzero Neutrino Masses with an Update on Dynamical Dark Energy and Lensing Anomaly}",
    eprint = "2504.15340",
    archivePrefix = "arXiv",
    primaryClass = "astro-ph.CO",
    doi = "10.3847/2041-8213/ade1cc",
    journal = "Astrophys. J. Lett.",
    volume = "986",
    number = "2",
    pages = "L31",
    year = "2025"
}

@article{Sabogal:2025jbo,
    author = "Sabogal, Miguel A. and Nunes, Rafael C.",
    title = "{Robust Evidence for Dynamical Dark Energy from DESI Galaxy-CMB Lensing Cross-Correlation and Geometric Probes}",
    eprint = "2505.24465",
    archivePrefix = "arXiv",
    primaryClass = "astro-ph.CO",
    month = "5",
    year = "2025"
}

@article{Silva:2025twg,
    author = "Silva, Emanuelly and Nunes, Rafael C.",
    title = "{Testing Signatures of Phantom Crossing through Full-Shape Galaxy Clustering Analysis}",
    eprint = "2507.13989",
    archivePrefix = "arXiv",
    primaryClass = "astro-ph.CO",
    month = "7",
    year = "2025"
}

@article{Khoury:2025txd,
    author = "Khoury, Justin and Lin, Meng-Xiang and Trodden, Mark",
    title = "{Apparent $w<-1$ and a Lower $S_8$ from Dark Axion and Dark Baryons Interactions}",
    eprint = "2503.16415",
    archivePrefix = "arXiv",
    primaryClass = "astro-ph.CO",
    month = "3",
    year = "2025"
}

@article{Cruz:2019kgz,
    author = "Cruz, Miguel and Lepe, Samuel and Morales-Navarrete, Gerardo",
    title = "{Qualitative description of the universe in the interacting fluids scheme}",
    eprint = "1902.09684",
    archivePrefix = "arXiv",
    primaryClass = "gr-qc",
    doi = "10.1016/j.nuclphysb.2019.114623",
    journal = "Nucl. Phys.",
    volume = "B",
    pages = "114623",
    year = "2019"
}

@article{Cardenas:2018nem,
    author = "C{\'a}rdenas, V{\'\i}ctor H. and Grand{\'o}n, Daniela and Lepe, Samuel",
    title = "{Dark energy and Dark matter interaction in light of the second law of thermodynamics}",
    eprint = "1812.03540",
    archivePrefix = "arXiv",
    primaryClass = "astro-ph.CO",
    doi = "10.1140/epjc/s10052-019-6887-0",
    journal = "Eur. Phys. J. C",
    volume = "79",
    number = "4",
    pages = "357",
    year = "2019"
}

@article{Cheng:2025lod,
    author = "Cheng, Hanyu and Di Valentino, Eleonora and Escamilla, Luis A. and Sen, Anjan A. and Visinelli, Luca",
    title = "{Pressure Parametrization of Dark Energy: First and Second-Order Constraints with Latest Cosmological Data}",
    eprint = "2505.02932",
    archivePrefix = "arXiv",
    primaryClass = "astro-ph.CO",
    month = "5",
    year = "2025"
}

@article{Afroz:2025iwo,
    author = "Afroz, Samsuzzaman and Mukherjee, Suvodip",
    title = "{Hint towards inconsistency between BAO and Supernovae Dataset: The Evidence of Redshift Evolving Dark Energy from DESI DR2 is Absent}",
    eprint = "2504.16868",
    archivePrefix = "arXiv",
    primaryClass = "astro-ph.CO",
    month = "4",
    year = "2025"
}

@article{Mishra:2025goj,
    author = "Mishra, Swagat S. and Matthewson, William L. and Sahni, Varun and Shafieloo, Arman and Shtanov, Yuri",
    title = "{Braneworld Dark Energy in light of DESI DR2}",
    eprint = "2507.07193",
    archivePrefix = "arXiv",
    primaryClass = "astro-ph.CO",
    month = "7",
    year = "2025"
}

@article{Trotta:2008qt,
    author = "Trotta, Roberto",
    title = "{Bayes in the sky: Bayesian inference and model selection in cosmology}",
    eprint = "0803.4089",
    archivePrefix = "arXiv",
    primaryClass = "astro-ph",
    doi = "10.1080/00107510802066753",
    journal = "Contemp. Phys.",
    volume = "49",
    pages = "71--104",
    year = "2008"
}

@article{Liddle:2007fy,
    author = "Liddle, Andrew R",
    title = "{Information criteria for astrophysical model selection}",
    eprint = "astro-ph/0701113",
    archivePrefix = "arXiv",
    doi = "10.1111/j.1745-3933.2007.00306.x",
    journal = "Mon. Not. Roy. Astron. Soc.",
    volume = "377",
    pages = "L74--L78",
    year = "2007"
}

@Inbook{Pearson1992,
author="Pearson, Karl",
editor="Kotz, Samuel
and Johnson, Norman L.",
title="On the Criterion that a Given System of Deviations from the Probable in the Case of a Correlated System of Variables is Such that it Can be Reasonably Supposed to have Arisen from Random Sampling",
bookTitle="Breakthroughs in Statistics: Methodology and Distribution",
year="1992",
publisher="Springer New York",
address="New York, NY",
pages="11--28",
isbn="978-1-4612-4380-9",
doi="10.1007/978-1-4612-4380-9_2",
url="https://doi.org/10.1007/978-1-4612-4380-9_2"
}

@article{Robert_2009,
   title={Harold Jeffreys’s Theory of Probability Revisited},
   volume={24},
   ISSN={0883-4237},
   url={http://dx.doi.org/10.1214/09-STS284},
   DOI={10.1214/09-sts284},
   number={2},
   journal={Statistical Science},
   publisher={Institute of Mathematical Statistics},
   author={Robert, Christian P. and Chopin, Nicolas and Rousseau, Judith},
   year={2009},
   month=may }

@article{KassRaftery1995,
  title        = {Bayes Factors},
  author       = {Kass, Robert E. and Raftery, Adrian E.},
  journal      = {Journal of the American Statistical Association},
  volume       = {90},
  number       = {430},
  pages        = {773--795},
  year         = {1995},
  publisher    = {Taylor \& Francis},
  doi          = {10.1080/01621459.1995.10476572},
  url          = {https://www.tandfonline.com/doi/abs/10.1080/01621459.1995.10476572}
}

@ARTICLE{1100705,
  author={Akaike, H.},
  journal={IEEE Transactions on Automatic Control}, 
  title={A new look at the statistical model identification}, 
  year={1974},
  volume={19},
  number={6},
  pages={716-723},
  doi={10.1109/TAC.1974.1100705}
}

@article{Schwarz1978,
  author       = {Schwarz, Gideon},
  title        = {Estimating the Dimension of a Model},
  journal      = {The Annals of Statistics},
  volume       = {6},
  number       = {2},
  pages        = {461--464},
  year         = {1978},
  url          = {http://www.jstor.org/stable/2958889}
}

@article{Spiegelhalter2002,
    author  = {Spiegelhalter, David J. and Best, Nicola G. and Carlin, Bradley P. and Van Der Linde, Angelika},
    title   = {Bayesian Measures of Model Complexity and Fit},
    journal = {Journal of the Royal Statistical Society: Series B (Statistical Methodology)},
    volume  = {64},
    number  = {4},
    pages   = {583--639},
    year    = {2002},
    issn    = {1369-7412},
    doi     = {10.1111/1467-9868.00353},
    url     = {https://doi.org/10.1111/1467-9868.00353},
    eprint  = {https://academic.oup.com/jrsssb/article-pdf/64/4/583/49723641/jrsssb\_64\_4\_583.pdf}
}

@article{Bedroya:2025fwh,
    author = "Bedroya, Alek and Obied, Georges and Vafa, Cumrun and Wu, David H.",
    title = "{Evolving Dark Sector and the Dark Dimension Scenario}",
    eprint = "2507.03090",
    archivePrefix = "arXiv",
    primaryClass = "astro-ph.CO",
    month = "7",
    year = "2025"
}

@article{Banerjee:2020xcn,
    author = "Banerjee, Aritra and Cai, Haiying and Heisenberg, Lavinia and Colg{\'a}in, Eoin {\'O}. and Sheikh-Jabbari, M. M. and Yang, Tao",
    title = "{Hubble sinks in the low-redshift swampland}",
    eprint = "2006.00244",
    archivePrefix = "arXiv",
    primaryClass = "astro-ph.CO",
    doi = "10.1103/PhysRevD.103.L081305",
    journal = "Phys. Rev. D",
    volume = "103",
    number = "8",
    pages = "L081305",
    year = "2021"
}

\end{document}